\documentclass{article}
\usepackage{cite}
\usepackage{amsmath,amssymb,amsfonts}
\usepackage{algorithmic}
\usepackage{graphicx}
\usepackage{textcomp}
\usepackage{xcolor}
\usepackage{graphicx}
\usepackage{cite}
\usepackage{picinpar}
\usepackage{amsmath}
\usepackage{url}
\usepackage{flushend}
\usepackage[latin1]{inputenc}
\usepackage{colortbl}
\usepackage{soul}
\usepackage{multirow}
\usepackage{makecell}
\usepackage{pifont}
\usepackage{xcolor}
\usepackage{alltt}
\usepackage[hidelinks]{hyperref}
\usepackage{enumerate}
\usepackage{siunitx}
\usepackage{breakurl}
\usepackage{epstopdf}
\usepackage{pbox}
\usepackage{amsfonts}
\usepackage{mathtools}
\usepackage{commath}
\usepackage{booktabs}
\usepackage{caption}
\usepackage{subcaption}
\usepackage{bm}
\usepackage{mathrsfs}

\def\BibTeX{{\rm B\kern-.05em{\sc i\kern-.025em b}\kern-.08em
    T\kern-.1667em\lower.7ex\hbox{E}\kern-.125emX}}

\usepackage{tikz}
\usetikzlibrary{spy}
\usepackage{circuitikz}
\usetikzlibrary{patterns,angles,quotes,intersections,calc}

\newcommand{\energy}{\Bar{\mathcal{E}}}
\newcommand{\denergy}{\dot{\bar{\mathcal{E}}}}
\newcommand{\Energy}{\bm{\Xi}}
\newcommand{\dEnergy}{\bm{\dot{\Xi}}}

\newcommand{\power}{P}
\newcommand{\Power}{\mathcal{P}}
\newcommand{\control}{\sigma}
\newcommand{\renergy}{\mathcal{E}^{\star}}
\newcommand{\ta}{T^a}
\newcommand{\td}{T^d}

\newcommand{\tcycle}{\mathcal{T}^{cy}}
\newcommand{\cost}{\mathcal{F}}

\newcommand{\gpowerlim}{P^{G}}

\newcommand{\switchset}{\bm{\gamma_{\sigma}}}
\newcommand{\noofchports}{n}

\newcommand{\Control}{\bm{\gamma_{\sigma}}}

\newcommand{\ecocost}{\psi}
\newcommand{\gpowerlimopt}{P^{G^{\star}}}
\newcommand{\pnooftimeslots}{n_p}
\newcommand{\mpower}{P^{\text{ch\_max}}}
\newcommand{\thr}{\delta_{\text{1}}}
\newcommand{\thrf}{\delta_{\text{2}}}
\newcommand{\ith}{i^{\text{th}}}
\newcommand{\eenergy}{\mathcal{E}}

\newcommand{\Eenergy}{\bm{E}}
\newcommand{\sspace}{\Omega_S}
\newcommand{\dspace}{\Omega_D}
\newcommand{\whole}{\mathbb{Z}_{\geq0}}
\newcommand{\expNa}{\hat{N}^a}
\newcommand{\expNat}{\hat{N}^a_{k}}
\newcommand{\expNap}{\hat{N}^a_{k+1}}
\newcommand{\expenergys}{{_{s}}\hat{\mathcal{E}}^{\star}}

\begin{document}

\title{A Two-Layers Predictive Algorithm \\ for Workplace EV Charging \\
\thanks{ %is with the Control Systems Technology group at the Eindhoven University of Technology, Netherlands. email:j.j.a.baltussen@student.tue.nl \\
Saif Ahmad, Jochem Baltussen, Pauline Kergus, Zohra Kader and St\'ephane Caux are with CODIASE Group at LAPLACE (Universit\'e de Toulouse, CNRS, INPT, UPS), Toulouse, France.
email: \{ahmad, kergus, kader, caux\}@laplace.univ-tlse.fr,j.j.a.baltussen@student.tue.nl\\This work is funded by ADEME, the French Agency for Ecological Transition, as part of the I-REVE project.}
}

\date{}
\author{S. Ahmad, J. Baltussen, P. Kergus, Z. Kader and S. Caux}

\maketitle

\begin{abstract}
In this paper, the problem of electric vehicle (EV) charging at the workplace is addressed via a two-layer predictive algorithm. We consider a time of use (TOU) pricing model for energy drawn from the grid and try to minimize the charging cost incurred by the EV charging station (EVCS) operator via an economic layer based on dynamic programming (DP) approach. An adaptive prediction algorithm based on a non-parametric stochastic model computes the projected EV load demand over the day which helps in the selection of optimal loading policy for the EVs in the economic layer. The second layer is a scheduling algorithm designed to share the allocated power limit (obtained from economic layer) among the charging EVs during each charge cycle. The modeling and validation is performed using ACN  data-set from Caltech. Comparison of the proposed scheme with a conventional DP algorithm illustrates its effectiveness in terms of supplying the requested energy despite lacking user input for departure time.

\end{abstract}


\section{Introduction}
Electric vehicles (EVs) offer a promising solution to curtail the devastating environmental impact of fossil fuels, particularly in the transportation sector. Over the past decade, EVs have witnessed a rapid increase in their market share worldwide with around 6.6 million units sold in 2021, a 100\% jump from the previous year  \cite{outlook2022entering}.  Indeed, the prolific rise in number of EVs can be attributed primarily towards rising environmental consciousness, enactment of governmental policies to curb the detrimental impact of fossil fuels as well as incentivise EV adoption, and steady improvement in energy storage technologies. However, efforts to increase the number of EV charging stations (EVCS) have not been able to keep pace with the rising EV units on road and therefore, inadequate charging infrastructure remains one of major hurdles impeding their rapid adoption \cite{Deloitte2020EVfor2030}. 

  A majority of EVCSs cannot scale up their charging capacities due to infrastructure constraints. Furthermore, most EV chargers deliver an average power of around 7 kW which is approximately equal to the maximum power draw of a average household (meter power capacity of around 70\% homes in France is 6kVA). This implies that a medium sized workplace EVCS with uncoordinated charging using conventional methods such as fist-come-first-served (FCFS) can draw power equivalent to an entire neighbourhood during peak hours which can adversely affect the grid stability \cite{acnThesis} and result in high electricity bills. The ability to delay EV charging, particularly at workplaces or public charging places \cite{ACN_description}, provides inherent flexibility that can be exploited to limit the EVCS power draw based on the limitations of electrical infrastructure and optimize against fluctuating electricity prices.  Additionally, this allows EVCS operators to oversubscribe critical (and expensive) electrical infrastructure so as to increase the number of charging ports (CPs) without incurring significant upgrade costs. Exploiting this flexibility, however, requires smart charging algorithms capable of handling uncertainties arising from EV energy demand, user behaviour and battery management system (BMS) associated with each EV. In line with this fact, a number of techniques have been proposed in literature for EV charging such as \cite{RR_faircharge,fair_2_alt_to_price,ACNdata,threeStepApproachDSM,KDE2018caltechConf}. Scheduling algorithms for fair distribution of available power were introduced in \cite{RR_faircharge, fair_2_alt_to_price} and performance was compared to conventional approaches such as FCFS, first depart first served (FDFS) and round-robin (RR) without any regulation of the total power drawn from the grid. User behavior was modeled based on Gaussian mixture models (GMMs) in \cite{ACNdata}, and the predictions were used for selecting optimal size of onsite solar generation as well as flattening the EV load profile over the day. A three-step solution to demand side management of plug-in hybrid EV was implemented in \cite{threeStepApproachDSM} where the optimization was performed using dynamic programming (DP) algorithm and the allocated power was scheduled based on a demand vector associated with each EV. A diffusion based kernel density estimator was used in \cite{KDE2018caltechConf} for predicting user behavior and the charge scheduling problem was solved in a distributed manner using alternating direction method of multipliers (ADMM) approach.

Scheduling of EV charging presents a number of challenges such as the need for the charging algorithm to run in real-time, being able to function with limited user input and predict future arrivals in order to optimally allocate the power capacity. Furthermore, most charging algorithms do not take into consideration the discrete nature of the control signals for the CPs or variables over which the optimization should be performed. i.e. the optimization variables do not take continuous signals and lie in a discrete set \cite{ReviewEVstateOftheArt}. In this work, we develop a two-layers predictive charging algorithm that computes optimal loading policy for the EVs in real-time. As a first step towards smart charging, we implement an adaptive prediction algorithm for expected EV arrivals and load demand over the day which acts as an input for the economic optimization layer. The predictions are obtained via a non-parametric stochastic model which does not consider an underlying distribution for the incoming data (in contrast to \cite{ACNdata}) and is adapted in real-time based on the arrivals. The economic layer optimizes the peak power draw from the grid by minimizing the energy cost (based on a time of use (TOU) pricing model), following which a scheduling algorithm is implement in a second layer to decide the state for each CP. We consider that the control signal for the CP has only two states (ON, OFF) that determine whether the battery of a connected EV gets charged or not in each charge cycle. This represents the simplest charging scenario and is the essence of time-scheduling in EV charging. However, the results can be generalised to have multiple discrete charging levels to allow for power scheduling, in order to represent a more realistic operating scenario \cite{acnLeeZach2020online}.  The dynamics of BMS is included in the charging model for the battery to simulate a more practical scenario. To analyse the effectiveness of developed algorithm, we use ACN database \cite{ACNdata} that contains data from level 2 EVCS.

Remaining sections in this paper are organised as follows: Section \ref{sec_problem_form} presents a description of EVCS, modeling of the EV charging and internal dynamics of the BMS. Different elements of the two-layers predictive charging algorithms are introduced in Section \ref{sec_main_algo}. Section \ref{sec_numerical_simulation} contains numerical simulations evaluating the performance of proposed scheme on ACN data set. The paper ends in Section \ref{sec_conclusion} with a summary of conclusions and future perspectives.

\subsubsection*{Notations} $\mathbb{R}_+$ is the set of positive real numbers, $\whole:=\{0,1,2,..\}$ is the set of all non-negative integers, $\mathcal{I}_N$ is a set of natural numbers defined as $\mathcal{I}_N:=\{1,2,\dots,N\}$, $\texttt{diag}(a_1,\dots,a_n)$ denotes a diagonal matrix with $a_1,\dots,a_n$ as diagonal elements and $\bm{x}^T$ denotes transpose of vector. Notation for the variables is selected as  ${_{\text{(time slot of the day)}}X_{\text{(time instant)}}^{\text{(variable)}}}.$% while $\lceil \cdot\rceil$ and $\lfloor \cdot\rfloor$ denote the ceiling and floor operations, respectively.

\section{System Description and Modeling}\label{sec_problem_form}
\label{sec:problem_formulation}
We consider  the scenario  of an EVCS having $\noofchports$ number of CPs, each having the same max charging capacity $(\mpower)$ and can either be switched ON or OFF in a particular charging cycle of time interval $\tcycle$ (10 minutes interval). We assume that the following information set is available for an EV connect to $i^{\text{th}}$ CP: $\Sigma:=\{{^i}\ta,{^i}\renergy,{^i}\td\}$, where $\renergy_i$= energy to be supplied, ${^i}\ta$= time of arrival and ${^i}\td$= duration of departure. Once a particular (ON/OFF) state is imposed for a CP, it remains in that state during the charge cycle unless supplied energy reaches the desired reference or the EV departs.

% \textcolor{red}{Pauline: $\Sigma:=\{\ta_i,\renergy_i,\td_i\}$}

% \zohra{I have a doubt about the fact that we consider that we know the energy to be supplied? it is very conservative as assumption. If we even don't know the state of charge of the vehicle how can we assume that we know the energy to be supplied. Please explain better this point} \saif{we consider it as a user input, the user demands the energy that needs to be supplied to their vehicle and we supply what is being asked by the user.}

\begin{figure}
    \centering
    \includegraphics[width=0.7\textwidth]{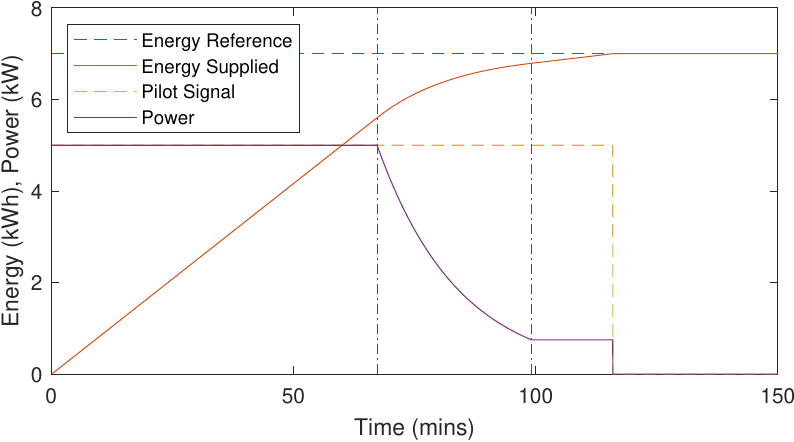}
    \caption{Variation in power supplied $(\power)$ with energy supplied $(\energy)$ where the vertical lines demarcate the three different regions during charging.}
    \label{fig:bms}
\end{figure}

The charging of EV battery is defined using normalized energy equation which is  of the form
\begin{equation}
{^i}\denergy(t)=\eta{^i}\power(t)/{^i}\renergy
\label{eq_energy_ode}
\end{equation}
% \zohra{Does $$\eta $$ has a physical unit because in the end the integral of $${^i}\denergy(t)$$ is an energy so the homogeneity of the equation must be ensured} \saif{$\eta$ is unitless constant, $\energy$ is the normalized energy equation and is thus unitless as well. }
where $\eta\in[0,1]$ is the efficiency of energy transfer (assumed constant and the same for each CP) while ${^i}\power\in\mathbb{R}_+$ is the power supplied. In a practical charging scenario, the pilot signal (power or current set-point for the CP, the former is considered in our work) only serves as an upper bound during the charging process and individual EVs charge as per the internal dynamics determined by their BMS. In order to simulate the uncertainties associated with practical charging, we consider a simplified dynamics of BMS which limits charging power as the SOC reaches a certain threshold. Since SOC data for the charging sessions is unavailable in the ACN data set, we define the power draw $\power_i$ in \eqref{eq_energy_ode} as a function of fractional energy supplied $({^i}\energy)$ instead which is given by
\begin{equation}
    {^i}\power=\begin{cases}
        & \mpower \ , \text{ if } \  {^i}\energy\leq\thr \ \\
        & \left( 1-{^i}\energy\right)\frac{\mpower}{1-\thr} \text{ , if } \thr<{^i}\energy\leq\thrf \\
        & \left(\frac{1-\thrf}{1-\thr}\right)\mpower \text{ , if } \thrf<{^i}\energy\leq1
    \end{cases}
    \label{eq_ev_bms}
\end{equation}
% \zohra{How can ${^i}\energy$ be compared to 1 or $\delta_1$ or $delta_2$ ? Something is messing in equation 2? May be the definition of ${^i}\energy$! the information about $\energy=7kWh$ is not sufficient. It even introduce more confusion } \saif{My mistake, it was a typo. ${^i}\renergy=7kWh$. Since ${^i}\energy$ is the normalised energy, it is unitless and can be compared to $\delta_1$ and $\delta_2$}

where $\thr$ and $\thrf$ are thresholds that determine the charging behavior. Measurements of power and energy supplied to battery are shown in Fig. \ref{fig:bms} where ${^i}\renergy=7kWh,\mpower=5kW, \thr=0.8,\thrf=0.97$ and $\eta=1$. It is to be noted that the pilot signal is made to coincide with completion of charging for the battery in Fig. \ref{fig:bms}, however, the power supplied goes to zero if the pilot signal becomes zero at any instance during charging which in turn is decided by the scheduling algorithm.

Let $\gpowerlim\in[\gpowerlim_{min},\gpowerlim_{max}]$ be the limit of power draw from the grid in a particular charge cycle where $\gpowerlim_{min}$ can be some local base load while $\gpowerlim_{max}$ is a hard limit imposed by electrical infrastructure of EVCS. It is assumed the power draw from the grid is a time dependent variable which can be optimized based on economic considerations and acts as an inequality constraint for the scheduling algorithm. The compact form of \eqref{eq_energy_ode} can be expressed as 
\begin{equation}
    \dEnergy(t)=\eta\Power(\sigma,t)
    \label{eq_Energy_ode_combined}
\end{equation}
where $\Energy:=[{^1}\energy,{^2}\energy,\dots,{^{\noofchports}}\energy]^T$ and $\Power(\control,t):=\texttt{diag}(\Control)\cdot[{^1}\power/{^1}\renergy,{^2}\power/{^2}\renergy,\dots,{^{\noofchports}}\power/{^{\noofchports}}\renergy]^T.$ Here $\switchset\in\{0,1\}^{1\times\noofchports}$ is the set of input variables which determines whether power is being supplied to the $\ith$ EV ($\switchset(i)=1$) or not ($\switchset(i)=0$) and $\sigma\in\mathcal{N}:=\{1,2,4,\dots,2^{\noofchports}\}$ is the mode which determines the selection of ON/OFF state for each CP. Each mode is associated with a particular configuration of the charging network with each switch having a fixed ON-OFF state.

% \saif{We can also define a vector $\alpha$ here which determines whether the charging port is active i.e. an EV is connected to it and the required energy is not delivered yet}.  

% \saif{describe the system from the perspective of a EVCS, where the charging ports form the switched system. This way it will be easier to define the dimensions. }
\subsubsection*{Remark 1} Extension to a generalised scenario having more number of discrete levels instead of simple ON/OFF control for CP would require the selection power supplied as $\min(P_{pilot},{^i}P)$ where $P^{pilot}$ is the pilot signal which is a fraction of the $\mpower$ and ${^i}P$ is obtained from the equation governing the BMS dynamics in \eqref{eq_ev_bms}. Also, the set of input signal $\Control$ would contain fractional values in accordance with the set of allowable pilot signals.

\section{Two-layers Predictive Charging Algorithm}\label{sec_main_algo}
Main objectives of the EV charging algorithm can be defined as follows:\newline
\textit{Objective 1:} Implementing a closed-loop controller managing the switching $(\Control)$ of the CPs in the EVCS such that $\renergy_i-\energy_i=0$ for each EV.\newline
\textit{Objective 2:} Meeting \textit{Objective 1} before departure time for each EV to ensure customer satisfaction.  \newline
\textit{Objective 3:} Regulating the power draw from the grid so as to minimise the charging cost incurred to the EVCS operator.

\begin{figure}
    \centering
    \includegraphics[width=0.7\textwidth]{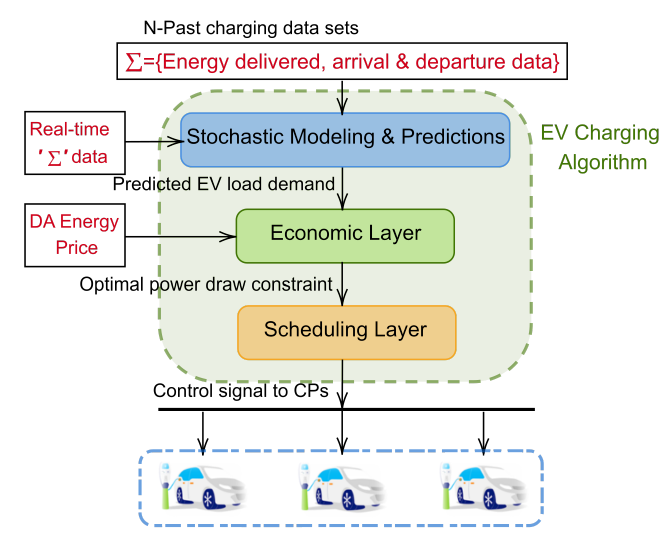}
    \caption{Block diagram of the predictive two-layers charging algorithm algorithm.}
    \label{fig:block_diag}
\end{figure}

In order to meet the aforementioned objectives, a two-layers predictive charging algorithm is introduced in this section which has three elements, (a.) predictions for the EV load, (b.) economic optimization to reduce operating cost and (c.) sharing of the allocated power capacity among charging EVs. These elements are explained in the following subsections. A block diagram of the algorithm is shown in Fig. \ref{fig:block_diag} showing the input variables (in red font), output from different blocks and interaction among various elements.
\label{sec:predictive_charging}
\subsection{Adaptive Prediction Algorithm} As mentioned earlier, a workplace charging scenario is considered in this paper. The charging data $\Sigma$ is obtained via ACN data set \cite{ACNdata}, from a site at the Caltech campus, containing 54 EV charging stations. It is assumed that the power supply to the EV stops as soon as energy delivered to the EV is equal to the  energy requested. Furthermore, the actual departure time of EVs can be different from those indicated by the user. 

% \jochem{This sentence is not true. We assume the delivered energy is equal to the requested energy (so we don't look at the user input for requested energy) and we take the actual departure time. Because there is a mismatch in indicated departure and actual departure this leaves some vehicles not loaded as much as they should, which is why we can compare this performance loss to our predictions. So we look at both the indicated departure and actual departure}

 In order to ensure faster adaptation in case of changes in the behavioural and usage patterns, past $500$ EV charging sessions are considered for modeling the user behavior. To accurately predict EV load demand at the EVCS, the expected number of arrivals per day $(\expNa)$ and their normalised distribution $(f^{N_a})$ is needed along with  energy demand for each EV given the arrival time. Furthermore, since the user input concerning departure time is mostly inaccurate \cite{ACNdata}, an estimate of expected departure time given the arrival time is also computed which reduces the number of user inputs required by the charging algorithm (a comparison of charging algorithms based on user input versus stochastic model for the departure time is presented in Section \ref{sec_numerical_simulation} to illustrate the benefit of considering the modeling approach).  $\expNa$ is computed via a moving average filter with a time window of 5 days wherein we distinguish between weekdays and weekends to account for the difference in arrival pattern. Furthermore, an estimate for $f^{N_a}$ is obtained via Kernel Density Estimator (KDE) from the discrete data \cite{EVmodelThesis} as
\begin{equation}
    \hat{f}^{N_a}(x)=\dfrac{1}{n_sh}\sum_{j=1}^{n_s}\mathcal{K}\left(\dfrac{x-x_j}{h}\right)
\end{equation}
where  $n_s=500$ is the number of discrete samples used for modelling, $x_j$ is the data sample, $\mathcal{K}(\cdot)$ is the Gaussian kernel function and $h$ is the bandwidth which determines the smoothness of estimated density. As mentioned in \cite{EVmodelThesis,KDE2018caltechConf}, the non-parametric density estimation approach does not assume the discrete data to have an underlying probability distribution which helps in more accurate modelling.

A statistical model for $\td$ and $\renergy$ is derived by splitting the day into $\pnooftimeslots=(24*60/\tcycle)$ number of slots, assigning each arrival to a time slot $s$ based on $\ta_i$ and then computing the estimated density of departure time $({_{s}}\hat{f}^{\td})$ and required energy $({_{s}}\hat{f}^{\renergy})$ for each time slot using KDE approach. The expected value of $\renergy$ obtained via KDE for a particular time slot $(\expenergys)$ is then used as the predicted value for all EVs with $\ta_i \in [(s-1)\tcycle, s\tcycle]$.

To account for prediction inaccuracy in terms of the number of arrivals as well as the estimated arrival Probability Distribution Function (PDF), we adapt the expected number of arrivals for the day as
\begin{equation}
    \expNap=[(C^{N_a}_{k}-\hat{C}^{N_a}_k)K^g_{k}+\expNat], k\in\{1,..,\pnooftimeslots\}
    \label{eq_adaptive_prediction_exp_Na}
\end{equation}
where $k$ is the discrete time step,
\begin{equation}
     K^g_k=\sqrt{(F^{N_a}_k+(k/\pnooftimeslots))/2}
     \label{eq_adaptation_gain}
\end{equation}is the adaptation gain, $F^{N_a}_k$ is the Cumulative Distribution Function (CDF) obtained from $\hat{f}^{N_a}$ while $C^{N_a}_k$ and $\hat{C}^{N_a}_k=\expNat \cdot F^{N_a}_k$ are the actual number of arrivals and the expected number of arrivals up to time instant $k$. The expression for $\expNat$ in \eqref{eq_adaptive_prediction_exp_Na} is inspired by the integral action in feedback control systems used for steering the output towards a desired reference which is the actual arrivals in our case. The adaptation gain $K^g_k$ is selected such that it is small towards the start of the day which results in slower adaptation of $\expNat$, and increases gradually based on the time of the day as well as $\hat{f}^{N_a}$. Furthermore, the square-root in expression \eqref{eq_adaptation_gain} results in a faster adaptation towards the middle of the day  compared to the linear case. It is worth mentioning that the expression in \eqref{eq_adaptation_gain} can be replaced by any piece-wise continuous function $f^n$ such that ${_{s=0}}f^n_{k=0}=0$ and ${_{s=\pnooftimeslots}}f^n_{k=\pnooftimeslots}=1$.  

The expected number of arrivals in a time slot $s$ at time interval $k$ are obtained as
\begin{equation}
    {_s}\expNat=\expNat\int_{(s-1)\tcycle}^{s\tcycle}\hat{f}^{N_a}(x)dx,
    \label{eq_exp_Na_per_time_slot}
\end{equation}
and is multiplied with $\expenergys$ to get the cumulative EV load demand.

 At each time $k$, a prediction set $\hat{\Sigma}$ containing the expected energy demand and departure PDF is computed for $s=\{k,k+1,\dots,\pnooftimeslots\}$. At the beginning of the day, the stochastic model is updated  using the data from past $500$ charging sessions and an initial prediction is made based on this historical data which is then updated at the end of each charge cycle.  

\subsection{Economic Layer}
The objective of the economic layer is to minimize an economic cost function based on the aggregated energy demand using dynamic programming (DP), similar to the approach in \cite{threeStepApproachDSM}. At first, the maximum and minimum energy constraint vector per EV is computed as:
\begin{equation}
\begin{split}
    {^i}\eenergy^{\max} = \{{^i}\eenergy^{\max}_k|{^i}\eenergy^{\max}_k = &\min(\mpower k\tcycle/60,{^i}\renergy)\},\\ &\forall k \in \{{^i}k_{a},...,{^i}k_{d}\}\\
    {^i}\eenergy^{\min} = \{{^i}\eenergy^{\min}_k|{^i}\eenergy^{\min}_k = &\max({^i}\renergy- \\ \mpower({^i}k_{d}-k)&\tcycle/60,0),
    \forall k \in \{{^i}k_{a},...,{^i}k_{d}\}
\end{split}
\label{eq: Emin}
\end{equation}
where ${^i}k_{a} \text{ and }{^i}k_{d}$ are the time slots associated with ${^i}\ta \text{ and }{^i}\td$, respectively, to obtain the aggregated constraints for the complete charging station
\begin{equation}
     \Eenergy^{\max} = {\sum^{n_{ev}}_{i=1}} {{^i}\eenergy^{\max}}, \ 
    \Eenergy^{\min} = {\sum^{n_{ev}}_{i=1}} {{^i}\eenergy^{\min}},
    \label{eq: Emax aggregated}
\end{equation}where $n_{ev}$ is the number of EVs. At any time $k$, the total energy supplied $(\eenergy^a_k)$ should be between $\Eenergy^{\max}_k$ and $\Eenergy^{\min}_k$, which forms the state space $\sspace$ for the optimization problem, so as to ensure that the energy demand of each EV is fulfilled before its departure time. If there are any essential (additional) loads that need to be supplied by the EVCS then the additional power drawn can be included to form the set $\sspace$, however, in this work we consider it to be zero i.e. $\gpowerlim_{min}=0$ at all times. Since we assume that the CPs can only be switched ON/OFF during a charge cycle, the amount of power drawn from the grid is an integer multiple of $\mpower$ which in turn determines the number of CPs that can be turned ON. The decision space $\dspace$ for the number of CPs to be turned ON $x_D$ depends on $\eenergy^a_k$ where
    $ \bigl \lceil\max(\Eenergy^{\min}_k-\eenergy^a_k,0)/\mpower\bigr \rceil\le x_D \le \bigl \lfloor\min(\Eenergy^{\max}_k-\eenergy^a_k,\gpowerlim_{max})/\mpower\bigr \rfloor$
and it is assumed that $\gpowerlim_{max}=\noofchports\mpower$  for the sake of simplicity. 
The objective for the economic layer can be cast as an optimization problem
\begin{equation}
    \begin{split}
        \min_{\gpowerlim}&\ecocost(\gpowerlim) \text{ subject to} \\
        &\eenergy^a \in \sspace, \ x_D \in \dspace
    \end{split}
\end{equation}
where $\ecocost(\cdot)$ takes the day-ahead (DA) hourly electricity cost $(E^{cost})$ as   input and generates the optimal power limit $\gpowerlimopt$ for each charge cycle.  In our case, we use the DP algorithm to compute optimal loading policy that recursively minimizes the cost according to the Bellman equation \cite{bellmanSource}
\begin{equation}
    J(\eenergy^a_k) = \min_{a_k}\{C(\eenergy^a_k,a_k)+J(\eenergy^a_{k+1})\}
    \label{eq_bellman}
\end{equation}
where action $a_k$ results in an associated cost $C$ in state $\eenergy^a_k$. The algorithm computes the optimal number of CPs that can be turned ON in each charge cycle based on the current energy state via backward induction method, which in turn acts as power draw constraint for the lower level scheduling task.

We consider the following two scenarios while computing the optimal loading policy for the EVs, which differ in the way they consider the departure time $(\td)$ associated with each EV: \newline
\textbf{[S1]}  In this scenario, only real-time data (no prediction) based on the arrivals $({^i}\ta)$ and user input for ${^i}\renergy$ as well as ${^i}\td$ is considered for computing the optimal loading policy and hence, the total energy to be supplied builds up with EV arrivals. Furthermore, the cost function $C$ in \eqref{eq_bellman} only depends on the action $a_k$ and denotes the cost of energy to move from one energy state to another based on $E^{cost}_k$.\newline
\textbf{[S2]} In the second scenario, the adaptive predictions $(\hat{\Sigma})$ obtained in the previous subsection are used for computing the optimal loading policy via DP algorithm. Instead of relying on user input for $\td$, expected departures are computed based on estimated PDF ${_s}\hat{f}^{\td}$ for actual as well as predicted arrivals. Furthermore, the expected energy demand for each predicted arrival is considered to compute the optimal loading policy during each charge cycle. The cost $C$ in this case has two parts, one that depends on the action $a_k$, similar to the previous scenario, while the other denotes a cost associated with each energy state in $\sspace$. The second part of the cost is computed based on how far $\eenergy^a_k$ is from $\Eenergy^{max}_k$ and the expected number of EVs that have departed by that time. If $\eenergy^a_k$ is $a$ units (of energy state) apart from $\Eenergy^{max}_k$ and $b$ number of vehicles are expected to have departed, then the cost associated with the particular state is computed as $abw$ where $w$ is a weight assigned to the second part of the cost. It is to be noted that in this scenario, the $\Eenergy^{min}$ line does not exist and we penalize the farther $\eenergy^a_k$ is from $\Eenergy^{max}$. It is to be noted that \textbf{S2} denotes the developed algorithm in this paper while \textbf{S1} (as well as \textbf{S3} and \textbf{S4} introduced later in Section \ref{sec_numerical_simulation}) are for comparison.

\subsubsection*{Remark 2}
It is important to note that the objectives defined at the beginning of the section rely heavily on the accuracy of the departure times of the EVs. If the departure time obtained either in the form of user input or predictions is inaccurate, the objectives will not be met. In particular, if EVs depart prior to their indicated or predicted departure times, \textit{Objectives 1} and \textit{3} will not be met. On the other hand, if EVs depart later, the cost incurred to the EVCS operator might be higher due to excess power drawn from the grid during a costly hour.
\subsection{Scheduling Layer}
The objective of scheduling algorithm is to allocate the available grid power $\gpowerlimopt$, determined by the economic layer in previous subsection,  among charging EVs. In order to ensure optimal scheduling of the EVs, we  minimize a cost function $\cost_k$ at each time step $k$ so as to select the appropriate state (ON/OFF) of each CP which can be expressed as
\begin{equation}
\begin{split}
        \min_{\Control}& \ \cost_k(\Control) \text{ subject to } \\
        % & \gpowerlimopt=\argmin_{\gpowerlim}\ecocost(\cdot)\\
        & \mathbf{1}_{1,n}\cdot\Power(\control)\leq \gpowerlimopt.
\end{split}
    \label{eq_optimization_problem}
\end{equation}
The cost function in the preceding optimization problem is defined as
\begin{equation}
    \cost_k(\Control)=-\Control \cdot F(k,\tau,e) ,
    \label{eq_cost_function}
\end{equation}
where $$F(k,\tau,e):=\left[\sum_{j={^1}k_a}^{k}{^1}\tau_{j}^{m_1}\cdot {^1}e_{j}^{m_2}, \dots, \sum_{j={^{\noofchports}}k_a}^{k}{^{\noofchports}}\tau_{j}^{m_1}\cdot {^{\noofchports}}e_{j}^{m_2}\right]^T$$ is a combination of time elapsed since arrival ${^i}\tau_j:=j-{^i}k_a$ and remaining energy to be supplied ${^i}e_j:={^i}\renergy(1-{^i}\energy_j)$ in the $j^{\text{th}}$ time interval, for  EV connected to $\ith$ CP, while ${^i}k_a,{^i}k_d$ are the same as in \eqref{eq: Emin} defined per CP. Furthermore, $m_1, m_2\in\whole$ are exponents that define the contribution of ${^i}e_j$ and ${^i}\tau_j$ in $F$, respectively and are considered as $m_1=m_2=1$ to have equal contribution of both the factors in cost function. It is worth mentioning that selection of $m_2=0$ essentially implements a FCFS algorithm without any regard to the remaining energy to be supplied to EV. The function $F$ resets the cost associated with particular EV as soon as it departs and initialises again upon a new connection. In addition, the cost is only computed for active CPs that have a connect EV with unfulfilled energy demand.

\section{Numerical Simulation} \label{sec_numerical_simulation}
In this section, the performance of the developed two-layers predictive charging algorithm is evaluated using a simulation study performed on the ACN data set. To have consistency, an AC level 2 charging station is considered similar to the one at Caltech having 54 CPs where each CP has $\mpower=7.36kW \ (230V\times32A/1000)$. The charging efficiency is considered to be $\eta=0.95$ for each CP while $\thr=0.8$ and $\thrf=0.95$ are considered for each EV to simulate the effect of BMS.
 
\begin{figure}
    \centering
    \includegraphics[width=0.7\textwidth]{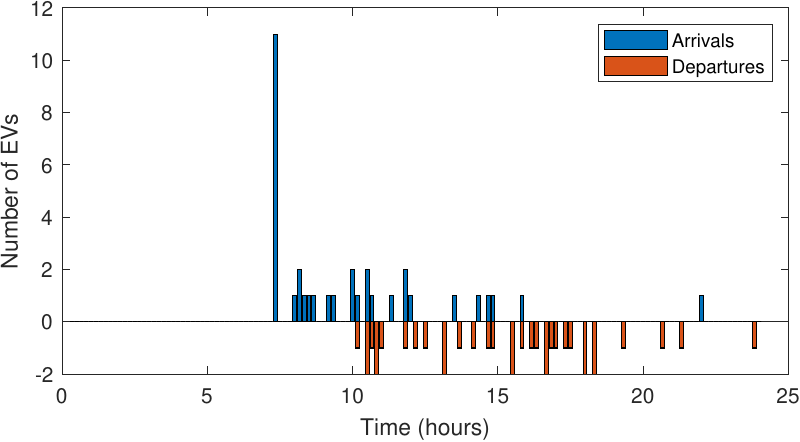}
    \caption{Arrival and departure pattern for day 32.}
    \label{fig: arrival departure pattern}
\end{figure} 
\begin{figure}
    \centering
    \includegraphics[width=0.7\textwidth]{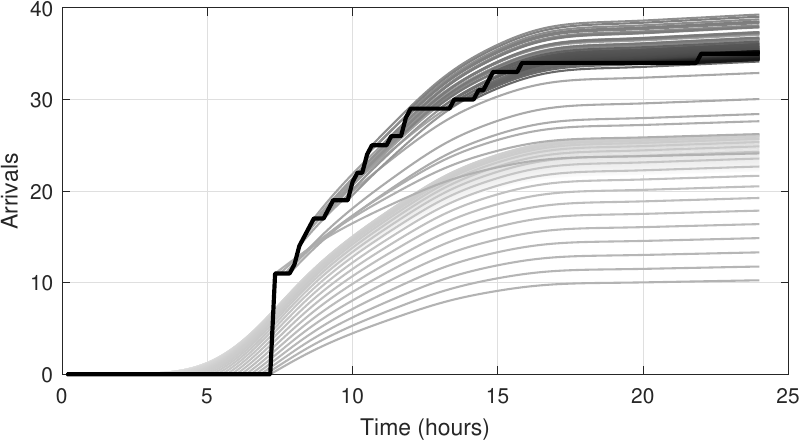}
    \caption{Predicted arrivals over the day with updates every 10 minutes (predictions are indicated by darker lines as we progress through the day with black line indicating actual arrivals).}
    \label{fig: Predicted arrivals}
\end{figure} 
%\begin{figure}
    %\centering
    %\includegraphics[width=0.49\textwidth]{Figures/Cumulative_plot.pdf}
    %\caption{Cumulative cost for 220 days and area under Emin curve}
    %\label{fig: cumulative cost}
%\end{figure}

 We consider EV data starting from $3^{\text{rd}}$ January 2021 and model the arrival, departure, and energy usage pattern with $\Sigma=\{{^i}\ta,{^i}\renergy,{^i}\td\}, i=\{1,2,\dots,500\}$ spread over 31 days and evaluate performance for day 32.  Furthermore, in order to illustrate the robustness of the developed algorithm towards uncertainty in user behavior (in terms of arrival and departure time) as well as consumption pattern (energy demand), we evaluate the performance of our algorithm over 220 days. The DA hourly energy price used in the economic layer is obtained from \cite{DynamicElectricityPricing} and remains the same for each day.  
 
The following test scenarios are considered in the simulation study. Scenarios \textbf{S1} and \textbf{S2} remain the same as defined in Section \ref{sec_main_algo} with $w=0.0003$ selected as the weight in \textbf{S2}. Scenario \textbf{S3} is the offline computation with perfect knowledge of arrival, departure, and energy demand throughout the day and serves as a benchmark for comparison. Scenario \textbf{S4} represents the conventional approach where EVs are charged on a first come first served basis without any limit on the power draw from the grid.

 The plot for EV arrivals and departures on day 32 is shown in Fig. \ref{fig: arrival departure pattern} and the predicted arrivals over the day is shown in Fig. \ref{fig: Predicted arrivals} where the expected arrivals per time slot are updated as the time progresses. At the start of the day, an initial prediction is made which decreases as per \eqref{eq_adaptive_prediction_exp_Na} and \eqref{eq_exp_Na_per_time_slot} when there are no arrivals during the morning hours and  is updated as soon as there is an uptick in arrivals. It is also evident from Fig. \ref{fig: Predicted arrivals} that the predictions become more accurate due to the adaptation as the day progresses. 
 \begin{figure}
    \centering
    \includegraphics[width=0.7\textwidth]{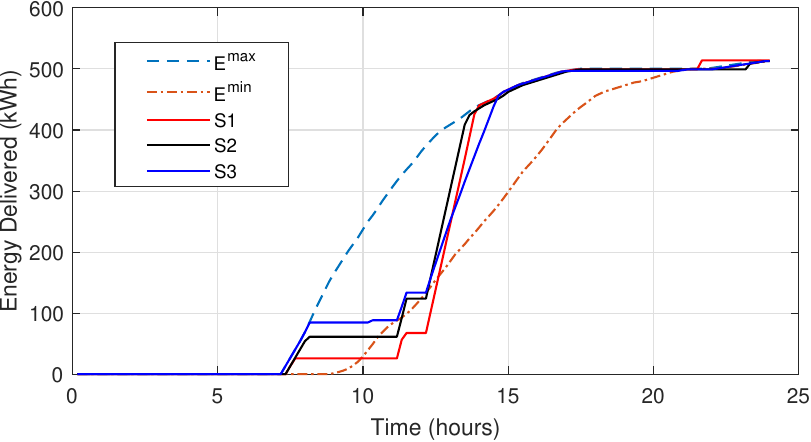}
    \caption{Energy supplied different DP strategies considered under \textbf{S1}, \textbf{S2} and \textbf{S3} for day 32.}
    \label{fig: Loading strategy}
\end{figure}
\begin{figure}
    \centering
    \includegraphics[width=0.7\textwidth]{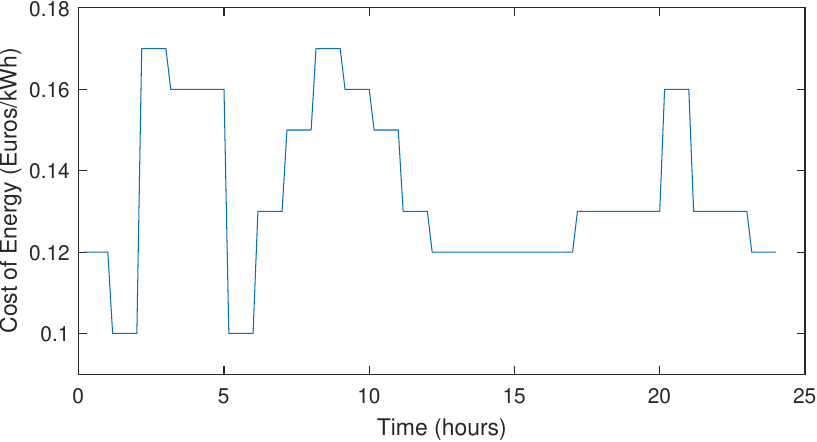}
    \caption{Variation in energy cost over the day.}
    \label{fig: energy cost}
\end{figure}

 \begin{figure}[]
     \centering
     \begin{subfigure}[b]{0.7\textwidth}
              \centering
         \includegraphics[width=\textwidth]{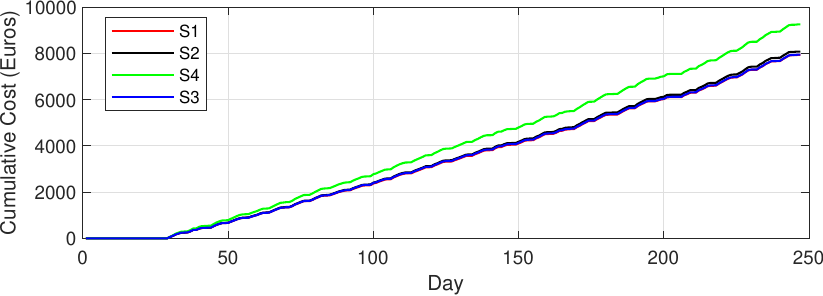}
         \caption{ Cost of Energy}
         \label{fig:cost of energy}
     \end{subfigure}
\\
       \begin{subfigure}[b]{0.7\textwidth}

     \centering
         \includegraphics[width=\textwidth]{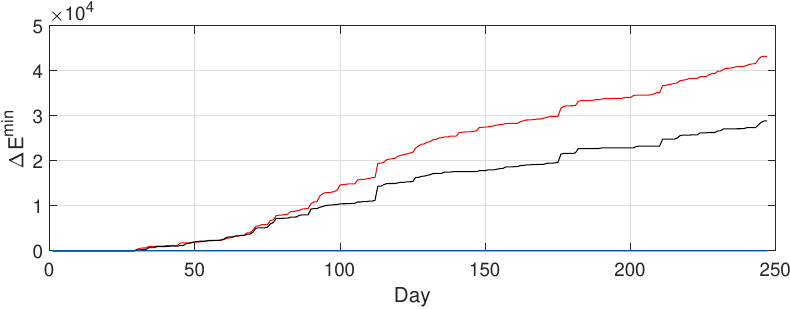}
         \caption{ $\Delta E^{min}$}
         \label{fig: area under emin}
     \end{subfigure}
        \caption{ Cumulative electricity cost and $\Delta E^{min}$ computed over 220 days on ACN data set.}
        \label{fig: cumulative 220}
\end{figure}
\begin{figure}
    \centering
    \includegraphics[width=0.7\textwidth]{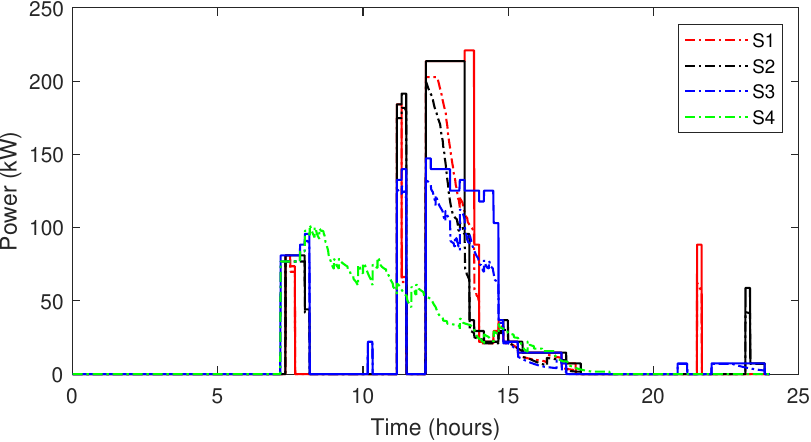}
    \caption{Allocated power limit $\gpowerlimopt$ (solid line) and actual power drawn (dash-dotted line).}
    \label{fig: scheduling layer}
\end{figure}

 The plot for optimal loading policy obtained in scenarios \textbf{S1}, \textbf{S2} and \textbf{S3} are shown in Fig. \ref{fig: Loading strategy} along with the cost of energy over the day in Fig. \ref{fig: energy cost}. The $\Eenergy^{max}$ and $\Eenergy^{min}$ lines in Fig. \ref{fig: Loading strategy} are the offline values computed at the end of the day. The loading policy in \textbf{S3} remains within $\sspace$ through the day which implies that under ideal charging conditions, the vehicles would be sufficiently charged when they depart. It is also evident that a sufficient number of CPs are not switched ON at the start of the day under \textbf{S1} and \textbf{S2} when there is an uptick in arrivals due to a mismatch between actual departure time and their expected value as well as user input. This is also evidenced by the loading policies going below $\Eenergy^{min}$ line which results in some vehicles departing without getting the requested energy. We compute the area between the $E^{min}$ curve and the $\eenergy^a$ as $\Delta E^{min}$ when $\eenergy^a<E^{min}$, which quantifies the expected number of EVs that depart with unfulfilled energy demands. The plot for $\Eenergy^{max}$ denotes the uncontrolled charging under \textbf{S4}. The plot for cumulative cost and $\Delta \Eenergy^{min}$ over 220 days obtained in Fig. \ref{fig: cumulative 220}  indicates that although the charging cost for all three scenarios is almost equivalent, $\Delta \Eenergy^{min}$ is significantly higher in \textbf{S1} compared to \textbf{S2} which indicates that more vehicles depart with unfulfilled energy demands under \textbf{S1}. It is also to be noted that the algorithm in \textbf{S2} can be made more reactive towards the price signal by reducing the weight $w$ in the cost function. As expected, the charging cost incurred in case of \textbf{S4} is significantly higher compared to \textbf{S1} and \textbf{S2}.

Plots for the scheduling layer under the considered scenarios are shown in Fig. \ref{fig: scheduling layer}. As mentioned earlier, the pilot signal for each EV acts as an upper bound for the power drawn and the EV charges as per the dynamics of its internal BMS. Hence the allocated power $\gpowerlimopt$ (shown by the solid lines) is not always utilized during the charging sessions and some of the EVs depart without charging to the desired level even under \textbf{S3} when perfect arrival and departure times are assumed to be known. Out of 500.39kWh requested by the EV users on day 32, the energy supplied under \textbf{S1, S2} and \textbf{S3} fell short by 111.8kWh (22.36\%), 111.18kWh (22.23\%) and 99kWh (19.8\%), respectively. Furthermore, 15, 15  and 18 vehicles were supplied the requested amount of energy while 20, 22 and 23 vehicle were supplied at least 90\% of the requested energy out of total 35 arrivals under \textbf{S1}, \textbf{S2} and \textbf{S3}. The results obtained for the scheduling layer corroborate the results obtained in Fig. \ref{fig: cumulative 220} in the sense that more number of vehicles are supplied the requested amount of energy with the proposed algorithm as compared to \textbf{S1}, despite the uncertainties associated with charging efficiency $(\eta)$ as well as BMS.

\section{Conclusion}\label{sec_conclusion}
\label{sec:conclusion}
A two-layers charging algorithm is developed in this paper for workplace EV charging stations. An adaptive prediction algorithm based on non-parametric stochastic model computes the expected arrivals, energy demand and departures over the day which acts as input to the economic layer. Comparison of the developed scheme with a DP algorithm that relies on user input for departure time illustrates that the predictive algorithm ensures better charging performance in terms of more number of EVs getting their demand fulfilled despite having similar cost. Furthermore, we also illustrate the effect of internal BMS while charging which further limits the energy supplied to the EVs under a constrained scenario. Results obtained from the  scheduling layer also shows that the proposed algorithm results in better optimal loading policies compared to the real-time DP algorithm despite the lack of user input for the departure time. A feedback of the mismatch in the allocated power level and actual power drawn from the grid, to the economic layer, will be investigated in future work to further improve charging performance of the developed algorithm. 

\bibliographystyle{IEEEtran}
\bibliography{biblio}   

\end{document}